\documentclass[floatfix,reprint,superscriptaddress,amsmath,amssymb,aps]{revtex4-1}
\usepackage[english]{babel}
\usepackage{csquotes}
\usepackage{subfiles}
\usepackage{xr} 

\usepackage[colorlinks=true]{hyperref}
\usepackage{cleveref}

\usepackage{graphicx} 
\usepackage{float} 
\usepackage[export]{adjustbox} 
\usepackage{tikz} 

\usepackage{makecell}
\makeatletter
\newcommand*{\addFileDependency}[1]{
  \typeout{(#1)}
  \@addtofilelist{#1}
  \IfFileExists{#1}{}{\typeout{No file #1.}}
}
\makeatother

\usepackage{upgreek}
\begin{document}
\title{An elongated quantum dot as a distributed charge sensor}
\newcommand\QMT{\,Quantum Motion, 9 Sterling Way, London N7 9HJ, United Kingdom}
\newcommand\UCL{\,London Centre for Nanotechnology, University College London, London WC1H 0AH, United Kingdom}
\newcommand\NBI{\,Center for Quantum Devices, Niels Bohr Institute, University of Copenhagen, Copenhagen, Denmark}
\newcommand\IMEC{\,imec, Kapeldreef 75, B-3001 Leuven, Belgium}
\author{S. M. Patom{\"a}ki}
\email{sofia@quantummotion.tech}
\affiliation{\QMT}
\affiliation{\UCL}

\author{J. Williams}
\affiliation{\QMT}
\affiliation{\UCL}
\author{F. Berritta}
\affiliation{\NBI}
\author{C. Lain{\'e}}
\affiliation{\QMT}
\affiliation{\UCL}
\author{M. A. Fogarty}
\affiliation{\QMT}
\author{R. C. C. Leon}
\affiliation{\QMT}
\author{J. Jussot} 
\affiliation{\IMEC}
\author{S. Kubicek} 
\affiliation{\IMEC}
\author{A. Chatterjee} 
\affiliation{\NBI}
\author{B. Govoreanu} 
\affiliation{\IMEC}
\author{F. Kuemmeth}
\affiliation{\NBI}
\author{J. J. L. Morton}
\email{john@quantummotion.tech}
\affiliation{\QMT}
\affiliation{\UCL}

\author{M. F. Gonzalez-Zalba}
\email{fernando@quantummotion.tech}
\affiliation{\QMT}

\date{\today}
\begin{abstract}
Increasing the separation between semiconductor quantum dots offers scaling advantages by facilitating gate routing and the integration of sensors and charge reservoirs. Elongated quantum dots have been utilized for this purpose in GaAs heterostructures to extend the range of spin-spin interactions. 
Here, we study a metal-oxide-semiconductor (MOS) device where two quantum dot arrays are separated by an elongated quantum dot (340~nm long, 50~nm wide). We monitor charge transitions of the elongated quantum dot by measuring radiofrequency single-electron currents to a reservoir to which we connect a lumped-element resonator. 
We operate the dot as a single electron box
to achieve charge sensing of remote quantum dots in each array, separated by a distance of 510~nm. Simultaneous charge detection on both ends of the elongated dot demonstrates that the charge is well distributed across its nominal length, supported by the simulated quantum-mechanical electron density. 
Our results illustrate how single-electron boxes can be realised with versatile footprints that may enable novel and compact quantum processor layouts, offering distributed charge sensing in addition to the possibility of mediated coupling. 
\end{abstract}
\maketitle
\section{Introduction}
In recent years, silicon spin qubits hosted in gate-defined quantum dots (QDs) have achieved major milestones making this platform a compelling option for large scale quantum computing~\cite{gonzalez2021scaling}. These include the demonstration of high fidelity one- and two-qubit gates on the same device~\cite{xue2022quantum, noiri2022fast, mills2022two}, high fidelity readout using radiofrequency (rf) single-electron transistors (SET)~\cite{connors2020rapid}, the demonstration of simple instances of quantum error correction~\cite{takeda2022quantum} and the scale up to 6-qubit devices in a linear arrangement~\cite{philips2022universal}. In addition, chips combining quantum and classical electronics  have been shown to operate at deep cryogenic temperatures, demonstrating a potential route for integrated addressing, control and measurement of qubits~\cite{guevel202019.2A, Ruffino2021a}.
\par
Silicon spin qubits typically rely on nearest neighbour exchange to implement two-qubit interactions~\cite{veldhorst2015two, zajac2017resonantly, huang2019fidelity}. Such a short-range qubit coupling applied across the qubit processor leads to high gate densities that hinder integration with local control electronics and gate fan-out~\cite{veldhorst2017silicon, boter2022spiderweb}, and introduce nonlinear responses due to cross-talk~\cite{undseth2022nonlinear}. Furthermore, introducing readout sensors within the qubit plane impacts the level of connectivity that can be achieved. To scale up beyond one-dimensional qubit arrays and integrate cryogenic electronics requires structures with enhanced functionality which can increase the separation between qubits, or between qubits and sensors. 
One approach to scaling is to use dispersive charge sensors, such as the rf single-electron box (SEB)~\cite{lafarge1991direct, house2016high, urdampilleta2019gate, cirianotejel2021spin}. The SEB offers similar levels of sensitivity to conventional charge sensors~\cite{oakes2022fast, niegemann2022parity} but only requires one charge reservoir, as opposed to two for the SET, facilitating the design of qubit arrays with higher connectivity. 
Another approach is to space out qubits by using elongated quantum dots (EQD) to mediate exchange interactions between them~\cite{martins2017negative, malinowski2018spin,wang2022jellybean}. Such an approach, requiring tunnel coupling between each of the remote QDs and the EQD, has been demonstrated in GaAs heterostructures to mediate fast, coherent exchange interaction between single spins separated by half a micron~\cite{malinowski2019fast}. A further advantage of the EQD is that it could itself act as a local charge reservoir to facilitate initialization~\cite{cai2019silicon}. 
\par
In this Article, we combine aspects of these two concepts to demonstrate an SEB with an elongated charge island that enables charge sensing of multiple remote QDs, which, due to the increased separation, show minimal cross-talk. The structure is fabricated using a three-layer $n^{+}$-doped polycrystalline silicon gate metal-oxide-semiconductor (MOS) process that enables the formation of the elongated SEB as well as few-electron QDs. The extended distribution and quantisation of the charge within the EQD, consistent with semi-classical modelling, allows it to sense the charge on QDs separated by over 0.5~$\mu$m. Finally, we show tunnel coupling between the remote QDs and the EQD, which fulfills one of the requirements for coherent mediated exchange.
\section{Experimental Methods} 
\label{sec:experimental_methods}
\begin{figure}
\includegraphics[]{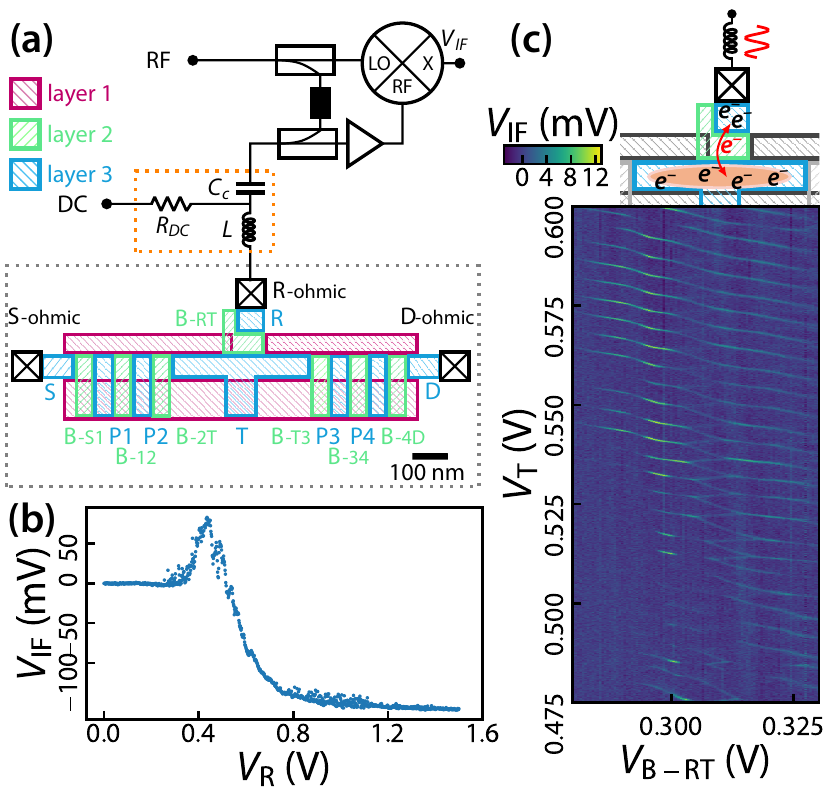}
\caption{
\label{fig:figure_1}
\textbf{Formation of an elongated single electron box.} 
\textbf{(a)} Device schematic (gray dotted rectangle) with simplified RF circuit diagram (signal filtering omitted). A lumped-element resonator (orange dotted rectangle) is galvanically attached to the ohmic contact below the accumulation gate R and monitored via changes in the demodulated baseband-frequency reflectometry signal, $V_{\mathrm{IF}}$. 
\textbf{(b)} 
Changes in $V_{\mathrm{IF}}$ reflect the accumulation of a 2DEG with increasing reservoir gate voltage $V_{\mathrm{R}}$. All other gates are held at zero bias.   
\textbf{(c)} 
The elongated QD is operated as a single electron box. Here, gates at zero bias are drawn in grayscale, while biased gates are drawn in colour. Orange blobs are cartoons indicating locations of QDs of interest. An elongated, multi-electron quantum dot forms under gate T and is tunnel coupled to a charge reservoir accumulated under gate R. Driving the resonator at its natural frequency drives cyclic electron tunnelling between the reservoir R, and the elongated quantum dot under gate T. 
The (T, B-RT) stability diagram obtained at $V_{\mathrm{R}} = 1.5$ V shows dot-to-reservoir transitions that become increasingly regular with increasing $V_{\mathrm{T}}$. The signal strength depends on $V_{\mathrm{B-RT}}$, since the barrier voltage modulates the EQD-reservoir tunnel rate. 
}
\end{figure}
Our device consists of two double quantum dots (DQDs) separated by an EQD, nominally 340~nm long and 50~nm wide. The measured device is fabricated with three 30~nm thick in-situ $n^{+}$ phosphorus-doped polycrystalline silicon gate layers formed with a wafer-level electron-beam patterning process. The Si substrate is separated from the first gate layer with a 8~nm thick thermally grown SiO$_{2}$, patterned on high-resistivity ($>3$~k$\Omega$) p-type Si wafer to minimise the density of oxide defects. Gate layers are electrically isolated from one another with a 5~nm thick 
blocking high-temperature deposited SiO$_{2}$~\cite{stuyck2021uniform}.
A schematic of the measured device is shown in Fig.~\ref{fig:figure_1}~\textbf{(a)}. We employ one layer of gates (closest to the silicon substrate) to provide confinement for the three possible current paths connecting ohmic contacts, around the active region of the device.  
A second layer of gates is used to form barriers between the EQD, the QDs and the reservoirs.
As seen in other MOS QD arrays~\cite{veldhorst2014an}, QDs can also be formed under these `barrier' gates in the second layer, depending on applied gate voltages. A third gate layer is used as plungers to control the occupation of the EQD, the QDs, and the extension of two-dimensional electron gases (2DEG) from under accumulation gates, denoted as reservoir (R), source (S), and drain (D), overlapping with corresponding ohmics, towards the active region of the device. 
\par
The device is cooled down in an Oxford Instruments Triton dilution refrigerator equipped with QDevil DACs, thermalizing filters and high-bandwidth sample holders~\cite{qdevil}.  At base temperature (25~mK) we confirm the functionality of the device with gate electrode leakage tests, followed by pinch-off and saturation voltage measurements (see Appendix~\ref{sec:cryogenic_device_characterization} for the preliminary device characterization protocol).
\par
We detect charge transitions between the EQD and the reservoir using rf reflectometry~\cite{vigneau2022probing}, via a lumped-element resonator attached to the ohmic contact of the accumulation gate R, as illustrated in the inset of Fig.~\ref{fig:figure_1}~\textbf{(c)}. Further details of the rf reflectometry setup and data acquisition are presented in Fig.~\ref{fig:supplementary_figure_reflectometry}~\textbf{(a)}. 
The rf voltage $V_{\mathrm{rf}}$ drives single-electron AC tunneling currents between the reservoir and the EQD when not in Coulomb blockade. Cyclic tunneling manifests as changes in the complex impedance of the device, modifying the resonant frequency and matching impedance of the lumped-element resonator~\cite{gonzalez2015probing}. Fig.~\ref{fig:supplementary_figure_reflectometry}~\textbf{(b)} shows the vector network analyzer response of the resonator with gate R biased off/on. We apply a signal with frequency close to that of the resonator and the reflected signal, which carries information of the complex impedance of the SEB, is amplified and mixed down to produce the DC signal $V_{\mathrm{IF}}$.
By monitoring shifts in the observed charge transitions, we operate the EQD as an SEB sensor which can simultaneously sense QDs formed near either of its ends.
\section{Results}
\subsection{Single-electron box tune-up}
\begin{figure*}
\includegraphics[]{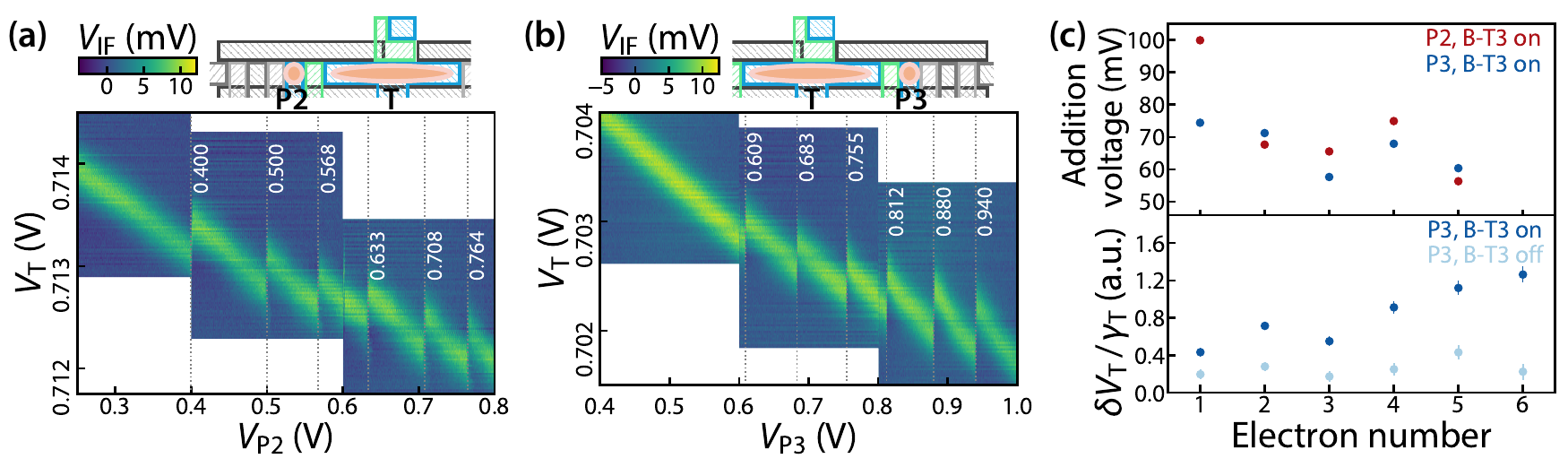}
\caption{
\label{fig:figure_2}
\textbf{Charge sensing of QDs under P2 and P3.} 
Operating point (top schematic) and discontinuities in the SEB peak locations (bottom dataset) reveal electron loading voltages for \textbf{(a)} P2 and \textbf{(b)} P3 (white numbers). 
\textbf{(c)} Upper panel shows the addition voltages extracted from \textbf{(a)}-\textbf{(b)}. Error bars, obtained from $V_{\mathrm{P2}}$ and $V_{\mathrm{P3}}$ resolution, are smaller than marker size.
\textbf{(c)} Lower panel shows the sensor peak shift, $\delta V_{T}$, with respect to peak linewidth, $\gamma_{T}$, at P3 QD charging events with B-T3 on (isolated from drain, as in panel \textbf{(b)}) and B-T3 off (connected to the reservoir formed with gate D). 
}
\end{figure*}
In order to operate the EQD as an SEB, we extend a 2DEG close to the active region of the device from a nearby ohmic contact by applying a positive voltage to gate R. We bias the EQD plunger gate, T, above the pinch-off voltage and tune the tunnel rate between the reservoir and the EQD by adjusting the voltage on the barrier gate B-RT. 
To tune the SEB, we first record $V_{\mathrm{IF}}$ as a function of $V_{\mathrm{R}}$ (see Fig.~\ref{fig:figure_1}~\textbf{(b)}). 
As $V_{\mathrm{R}}$ is increased, $V_{\mathrm{IF}}$ changes as the 2DEG is formed, modifying the circuit impedance. 
For $V_{\mathrm{R}} \gtrsim 1$~V, $V_{\mathrm{IF}}$ is nearly constant, indicating that the 2DEG is fully accumulated. In this region, changes in the resonator response due to voltage sweeps on the other gates can be ascribed to AC charge transport between 
the QDs and the 2DEG in the reservoir.
\par
Having fixed $V_{\mathrm{R}} = 1.5$~V, we then map out the charge stability diagram between gates T and B-RT (Fig.~\ref{fig:figure_1}~\textbf{(c)}), which shows dot-to-reservoir transitions (DRTs) indicating the presence of discretized charge states. For $V_{\mathrm{T}} \lesssim 0.55$~V, the data suggest a complex system comprising at least two coupled QDs, while for $V_{\mathrm{T}} \gtrsim 0.55$~V, the stability diagram increasingly resembles that of a single QD. Selecting $V_{\mathrm{B-RT}} = 0.29...0.31$~V maximizes the signal $V_{\mathrm{IF}}$ due to optimal tunnel rates between the reservoir and EQD. In the following, we use $V_{\mathrm{T}} = 0.69...0.72$~V, which we show to be sufficient for the EQD to extend over the length of the gate T. 
\subsection{Charge sensing of quantum dots}
We next use the EQD as an SEB to individually sense electrons in QDs under P2 and P3, and also as a local electron reservoir for these dots
(see Fig.~\ref{fig:figure_2}~\textbf{(a)}-\textbf{(b)}). 
To this end, starting from the SEB operating point of $V_{\mathrm{R}} = 1.5$~V, $V_{\mathrm{B-RT}} = 0.29$~V, and $V_{\mathrm{T}} = 0.70...0.72$~V, we further set $V_{\mathrm{B-2T}} = 0.250$~V, and $V_{\mathrm{B-T3}} = 0.225$~V. We illustrate this operating point with device schematics in Fig.~\ref{fig:figure_2}~\textbf{(a)} and \textbf{(b)}.
Positive barrier gate voltages increase tunnel rates from P2 to T and T to P3. A simulation of electron densities qualitatively illustrates how the barrier gates reshape and pull the QDs towards them. This effect is further discussed in Sec.~\ref{sec:simulated_quantum_mechanical_electron_densities}.  
Barrier gate voltages are chosen to reside below their observed first electron loading voltages, based on (B-2T,T) and (T,B-T3) stability diagrams (see Figs.~\ref{fig:supplementary_figure_barrier_sensing}~\textbf{(a)}-\textbf{(b)}).
\par
We detect the loading of an electron to either P2 or P3 QDs as a discontinuity in the SEB DRT, caused by the mutual capacitance between the EQD and the QDs. We mark the $0\to 1$ charge transitions as the first detected discontinuity. 
We find the first electrons to load at $V_{\mathrm{P2}}(0 \to 1) = 0.400$~V, and $V_{\mathrm{P3}}(0 \to 1) = 0.609$~V, respectively. Subsequent electrons load in steps of tens of millivolts. At occupancy of one electron, we find typical sensor peak voltage signal-to-noise ratios (SNR) of $\mathrm{SNR}_{\mathrm{P2}} = 10.7$ and $\mathrm{SNR}_{\mathrm{P3}} =14.6$, using an integration time of $1$~ms (see Appendix~\ref{sec:radiofrequency_reflectometry} for details).
\par
In order to understand whether the sensed QDs P2 and P3 are in the few-electron regime~\cite{lim2011spin}, we plot the extracted addition voltages in Fig.~\ref{fig:figure_2}~\textbf{(c)}. These addition voltages carry information of the electron-number-dependent confinement energies, as $V_{g}(n_{d} \to n_{d}+1) - V_{g}(n_{d}-1 \to n_{d}) = \alpha_{dg}^{-1} \big[ E_{C\,d}(n_{d}) + \Delta(n_{d}) \big]$, where $n_{d}$ is the electron number at the QD $d$; $\alpha_{dg}$ is the lever arm from QD $d$ to gate $g$; and $E_{C\,d}(n_{d}) + \Delta (n_{d})$ is the sum of the corresponding on-site charging energy and the confinement energy. The addition voltages are irregular in general and, in particular, we observe  an increase in the addition voltage both for P2 and P3 when loading from the presumed $4 \to 5$ electron state. This is consistent with filling the lowest two $\pm z$ valley-orbit states, such that the next electron occupies a higher-energy orbital state. 
\par
Using an estimated T addition voltage of $|e|^{-1} \alpha_{T\,T}^{-1} \, E_{C\,T} = 4.4 \pm 0.2$~mV (see Fig.~\ref{fig:supplementary_figure_Tgate_charging_energies}), 
loading the first electron under P2 and P3 induces a charge of 
$\mathrm{d}q = 0.075\,e \pm 0.01$ for P2, and 
$\mathrm{d}q = 0.032\,e \pm 0.01\,e$ for P3, respectively,
onto the SEB. 
We also show in Fig.~\ref{fig:figure_2}~\textbf{(c)} the shifts in $V_{\mathrm{T}}$ induced by P3 electron loading, $\delta V_{\mathrm{T}}$, relative to the fitted linewidth of the SEB DRT, $\gamma_{\mathrm{T}}$. This ratio $\delta V_{\mathrm{T}}/\gamma_{\mathrm{T}}$ is a proxy for charge sensitivity, and indicates whether the sensor is in the small or large signal regime~\cite{keith2019single}. When loading from the EQD, with $V_{\mathrm{B-T3}} = 0.225$~V, the shifts become larger than the line width of the sensor peak, i.e.\ $\delta V_{\mathrm{T}} \geq \gamma_{\mathrm{T}}$, by the fifth electron. We retain some sensitivity to the QDs even when the barrier gates to the EQD are off at zero bias. In this case, we resort to loading electrons under P3 from a reservoir formed via D. Here, we set $V_{\mathrm{B-T3}} = 0$~V, $V_{\mathrm{B-34}} = 0.275$~V,  $V_{\mathrm{P4}}$ and $V_{\mathrm{B-4D}}$ to $0.9$~V, and $V_{\mathrm{D}}$ to $1.5$~V. We note that the first electron under P3 at this operating point is found at $V_{\mathrm{P3}} = 0.387$~V. We find that in this operating point, the sensitivity is lower and increases more slowly. 
\subsection{Charge sensing coupled quantum dots}
\begin{figure*}
\includegraphics[]{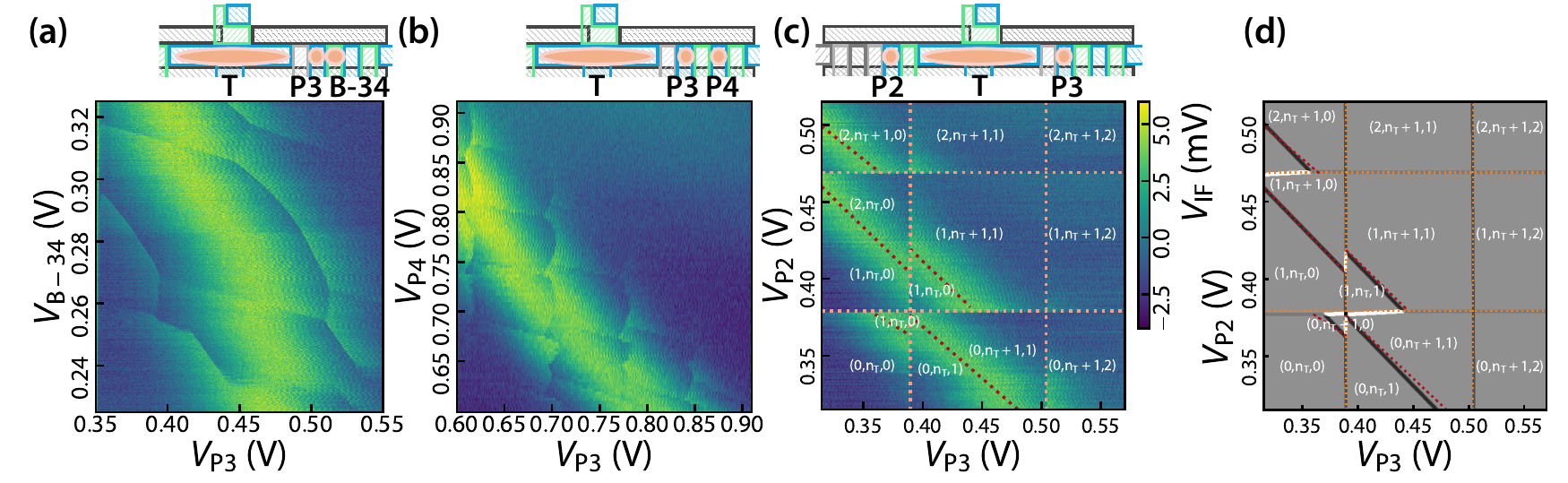}
\caption{
\label{fig:figure_3}
\textbf{Elongated single-electron-box as a distributed sensor.} 
\textbf{(a)}-\textbf{(c)} SEB charge-sensed stability diagrams of DQDs controlled with gates \textbf{(a)} P3 and B-34, \textbf{(b)} P3 and P4, and \textbf{(c)} TQD controlled with gates P2, T, and P3. Gate biasing and QDs are sketched with device schematics above the colour maps. 
\textbf{(a)} To define a DQD under P3 and B-34, we extend a 2DEG from the reservoir formed under gate D. We bias B-4D in saturation, and P4 near its pinch-off.
\textbf{(b)} To define a DQD under P3 and P4, we instead bias B-34 and B-4D as barriers. 
\textbf{(c)} To define a TQD between P2, T, and P3, we bias B-2T, B-T3, and B-34 as barriers. We bias $V_{\mathrm{T}} = 0.7093$ V to obtain a signal near the first P2 and P3 QD electrons. 
The estimated P2, T, and P3 QD charge occupations are indicated as $(n_{\mathrm{P2}}, n_{\mathrm{T}}, n_{\mathrm{P3}})$. 
\textbf{(d)} Grayscale colormap shows the voltage-cross-derivative of ground state of an electrostatic Hamiltonian, obtained using the experimentally estimated lever arms and charging energies. Orange and red dotted lines correspond to the fitted lines from panel \textbf{(c)}.}
\end{figure*}
Having established the basic operation of the EQD as a SEB charge sensor for nearby QDs, we next demonstrate its ability to sense different configurations of nearby coupled QDs. We then go on to assess the sensitivity of this distributed charge sensor with increasing distance. 
First, we form a DQD under P3 and B-34 by extending the reservoir 2DEG formed with gate D, setting $V_{\mathrm{B-4D}} = V_{\mathrm{P4}} = 0.9$~V, well above their threshold voltages, while operating P3 and B-34 close to their expected first electron voltages. We re-tune $V_{\mathrm{T}} = 0.7084$~V, retaining $V_{\mathrm{P3}}$ and $V_{\mathrm{B-34}}$ at the center of their selected voltage ranges. 
The resulting SEB-sensed (P3,B-34) stability diagram is shown in Fig.~\ref{fig:figure_3}~\textbf{(a)}. We observe a honeycomb pattern typical for a tunnel-coupled DQD, retaining sensitivity to charge transitions of both QDs, even though the center-to-center distance of the furthest dot to the EQD is $305$~nm. We measure local addition voltages of approximately $114 \pm 1$~mV and $43 \pm 1$~mV for P3 and B-34, respectively. 
\par
Second, we form a DQD under P3 and P4 (see Fig.~\ref{fig:figure_3}~\textbf{(b)}). Continuing from the previous operating point, we adjust the barrier voltages  $V_{\mathrm{B-4D}} = V_{\mathrm{B-34}} = 0.275$~V, while retaining $V_{\mathrm{B-T3}} = 0$~V, to create confinement, and retune $V_{\mathrm{T}} = 0.7068$~V.
Here, the DQD honeycomb pattern has average addition voltages of approximately $77 \pm 5$ and $63 \pm 5$~mV for P3 and P4, respectively. The observation of latching~\cite{yang2014charge}, i.e.\ distortion of P3 charge transitions, suggest that P3-P4 or P4-D tunnel rates are of the order of the ramp frequency $f_{\mathrm{ramp}}$ (see Appendix~\ref{sec:radiofrequency_reflectometry} for details on data acquisition). 
The center-to-center distance of P4 to the EQD is nominally $355$~nm, showing the charge sensing range of this extended SEB goes beyond those typically demonstrated by more conventional SEBs or   SETs~\cite{philips2022universal}. 
\par 
Finally, we form a triple quantum dot between P2, T, and P3, by drawing in electrons under P3 from the reservoir D, and under P2 from the EQD. We control tunnel rates to electron reservoirs with $V_{\mathrm{B-2T}} = 0.25$~V, $V_{\mathrm{B-T3}} = 0$~V, and $V_{\mathrm{B-34}} = 0.275$~V. 
We bias the SEB to $V_{\mathrm{T}} = 0.7093$~V, to maximise sensitivity when $V_{\mathrm{P2}}$ and $V_{\mathrm{P3}}$ are set close to their expected first electron voltages and
Fig.~\ref{fig:figure_3}~\textbf{(c)} shows the resulting (P2,P3) charge stability diagram of the triple QD. We label the estimated charge configuration for the P2, T, and P3 system as $(n_{\mathrm{P2}}, n_{\mathrm{T}}, n_{\mathrm{P3}})$. 
The estimates are based on a stability diagram simulation shown in Fig.~\ref{fig:figure_3}~\textbf{(d)}, which utilizes experimentally estimated lever arms and charging energies, which are further discussed in Sec.~\ref{sec:simulated_quantum_mechanical_electron_densities} and Appendix~\ref{sec:lever_arm_estimation}. 
The operating point is close to a so-called hextuple point, characterized by the hourglass shape, formed between $(0,n_{\mathrm{T}}+1,0)$ and $(1,n_{\mathrm{T}},1)$ charge states~\cite{fedele2020Thesis}. 
\par 
To confirm our understanding of the locations of the QDs in the triple QD configuration above, we extract the various lever arm ratios from the slope of the SEB peak and the quasi vertical and horizontal charge sensing shifts, obtained by line fits to the SEB peak positions (see Appendix~\ref{sec:lever_arm_estimation}). We observe close to zero P2-P3 cross-talk, as expected for remote QDs, with the estimate 
$\alpha_{\mathrm{P3},\mathrm{P2}}/\alpha_{\mathrm{P3},\mathrm{P3}} = (8 \pm 6) \times 10^{-3}$, 
obtained from the P3 charge transitions as a function of $V_{\mathrm{P3}}$. We get $\alpha_{\mathrm{P2},\mathrm{P3}}/\alpha_{\mathrm{P2},\mathrm{P2}} = 0 \pm [0, 3.33 \times 10^{-3}]$, limited by the lower data resolution along the $V_{\mathrm{P2}}$ axis. The average of the fitted EQD DRT slopes, marked with dashed dark red lines, is 
$\alpha_{\mathrm{T},\mathrm{P3}}/\alpha_{\mathrm{T},\mathrm{P2}} = 0.65 \pm 0.11$. 
A ratio equal to $1$ would indicate an EQD wavefunction which is symmetric with respect to locations of gates P2 and P3. Intuitively, the positively biased barrier B-2T ($V_{\mathrm{B-T3}} = 0$ V) pulls the EQD electron wavefunction towards P2, which could explain the lever arm asymmetry.
\par
Overall, the data from Fig.~\ref{fig:figure_3}~\textbf{(c)} demonstrates the simultaneous readout of QDs that are separated by approximately $510$~nm, operating the elongated SEB as a distributed charge sensor. 
The fact that a single EQD charge transition is capacitively shifted by the addition of charges to either P2 or P3 demonstrates that the EQD extends approximately over the length of gate T. 
We did not assess P2-T and T-P3 tunnel couplings at this operating point, however, in Appendix~\ref{sec:barrier_quantum_dot_sensing} we demonstrate that by utilizing dots under B-2T and B-T3 rather than P2 and P3, tunnel coupling to the EQD can be achieved. Our results demonstrate extended EQD wavefunctions and tunnel coupling to QDs in the periphery, both necessary requirements to utilize the EQD states for mediated exchange~\cite{srinivasa2015tunable}. 
\par
\subsection{Simulated quantum-mechanical electron densities}
\label{sec:simulated_quantum_mechanical_electron_densities}
\begin{figure}
\includegraphics[]{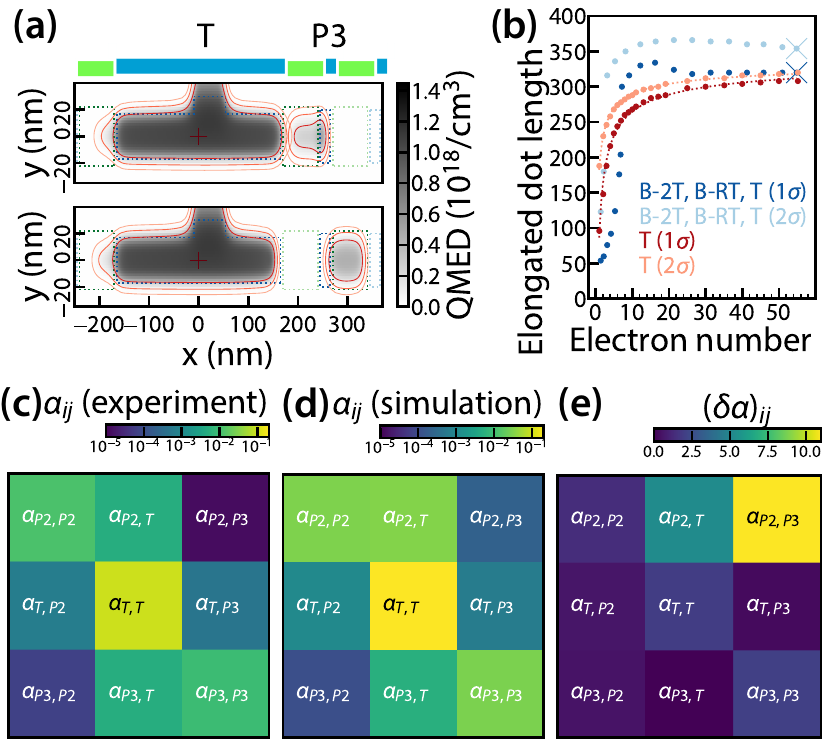}
\caption{
\label{fig:figure_4}
\textbf{Estimating the EQD length and the lever arm matrix.}
\textbf{(a)} Simulated QMEDs of the T-P3 DQD with B-T3 biased with a positive voltage (top panel; see Fig.~\ref{fig:figure_2} \textbf{(b)}) and at zero bias (bottom panel; Fig.~\ref{fig:figure_2} \textbf{(c)}) are shown as grayscale colormaps overlayed with layer 2 (green) and 3 (blue) gate locations (dotted rectangles). Red contours correspond to t $(1 - m \sigma) \rho_{\mathrm{max}}$ for $m = 1,2,3$. Gate side view (top) highlights the locations of gates T and B-T3. 
\textbf{(b)} EQD length as a function of electron numbers $n_{\mathrm{T}}$, integrated from the QMED. 
Red datasets are obtained by only biasing the gate T, and correspond to $(1 - m\sigma) \rho_{\mathrm{max}}$ for $m=1,2$ in increasing lightness. 
Dotted lines are fits to the power law $a n_{\mathrm{T}}^{-1/2} + b$. 
Blue datasets are obtained by biasing $V_{\mathrm{B-2T}} = 0.275$ V, $V_{\mathrm{B-RT}} = 0.3$ V, and varying $V_{\mathrm{T}}$, likewise $m=1,2$ are shown in increasing lightness. 
The cross markers correspond to the operating point of Fig.~\ref{fig:figure_3}~\textbf{(c)}. 
\textbf{(c)} Experimentally estimated lever arm matrix components. We use data from Figs.~\ref{fig:figure_2}-\ref{fig:figure_3}, together with an independent estimate for $\alpha_{T,T}$ to estimate the lever arm matrix. 
\textbf{(d)} Simulated lever arm matrix. 
Simulations use gate biases corresponding to experimental operating points, with each QMED corresponding to a QD simulated separately with up to nearest-neighbour gate biases. 
\textbf{(e)} Relative errors between experimentally estimated and simulated lever arm matrix components. 
}
\end{figure}
\par
To support the interpretation of a delocalized charge state under the EQD, and to benchmark our quantitative understanding of the QD systems under study, we employ a self-consistent Schr{\"o}dinger-Poisson solver (SPS) from a three-dimensional nanostructure simulation software~\cite{birner2011modeling, nextnanoWebManual} to evaluate so-called quantum-mechanical electron densities (QMED), denoted with $\rho(\textbf{r})$. We assimilate the QMEDs to probability densities under QDs to estimate shapes of many-electron charge states (see Appendix~\ref{sec:self_consistent_schroedinger_poisson_simulation} for details of the simulation methods). Figure~\ref{fig:figure_4}~\textbf{(a)} shows $(x,y)$ plane views of the simulated QMEDs of the T-P3 system studied in Fig.~\ref{fig:figure_2}~\textbf{(b)}-\textbf{(c)}.  
The two QMEDs are obtained by biasing the QD plunger gates (T or P3), and nearest neighbour barrier voltages at the non-zero biases where experimental data was taken. In the simulations, the barriers modify the shapes of the QDs, pulling QDs controlled with plunger gates towards the biased barriers, and extending the shape of the EQD. As we discuss below, the QD shape and location has an impact on (e.g.) lever arms, which are also experimentally measurable.
\par
The EQD length, obtained from the simulated $1\sigma$ and $2 \sigma$ QMED contours, is studied for a range of electron numbers, determined by integrating the simulated electron densities for a range of $V_{\mathrm{T}}$ voltages. The results are shown in Fig.~\ref{fig:figure_4}~\textbf{(b)}. 
In a simulation where only the gate T is biased, the EQD length increases monotonically. The EQD length can be fitted to  
the power law $x_{\mathrm{EQD}} = a n_{\mathrm{T}}^{-1/2} + b$, 
where $n_{\mathrm{T}}$ is the simulated electron number,
$a < 0$, and we find $b = 347$~nm and $b = 339$~nm for $1 \sigma$ and $2 \sigma$, respectively.  
\par
When B-2T and B-RT are also positively biased with constant voltages, the electron density under B-RT only, $n_{\mathrm{B-RT}} \approx 18.8$, is subtracted from the electron numbers. Here, the EQD length is a more complicated function of the electron number: The more graduate increase at low occupancy is due to how the B-RT gate pulls electrons, and the sharper increase at $n_{\mathrm{T}} \approx 6$ is caused by the EQD density merging with the density under B-2T.  As the electron number increases further, the EQD length (defined by $1\sigma$ or $2\sigma$) gradually decreases due to an increasing concentration of charge in the centre of the QD.
The simulated datapoints with $V_{\mathrm{T}} = 0.7093$~V (corresponding to the setpoint from Fig.~\ref{fig:figure_3}~\textbf{(c)}). 
The estimated length at this datapoint is $x = 320 \pm 2$~nm at $1 \sigma$, and $x = 354 \pm 2$~nm at $2 \sigma$. 
We use four measured datasets to estimate the lever arm components of the (P2,T,P3) system and compare them with simulated values, in Fig.~\ref{fig:figure_4}~\textbf{(c)}-\textbf{(e)}. Details of lever arm extraction, as well as all estimated and simulated lever arm components, are found in Appendices~\ref{sec:lever_arm_estimation},~\ref{sec:self_consistent_schroedinger_poisson_simulation}, and~\ref{sec:simulated_capacitance_matrices}. Simulated lever arms are systematically larger compared to experimentally extracted values, albeit typically agreeing within an order of magnitude. We find the largest errors for $\alpha_{\mathrm{P2,P3}}$ and $\alpha_{\mathrm{P2,T}}$ ($14.2$ and $5.3$, respectively), while the remaining off-diagonal lever arms have the smallest errors, from $0.074$ to $0.67$. 
\par
We simulate the TQD charge stability diagram from Fig.~\ref{fig:figure_3}~\textbf{(c)} using the estimated lever arm components from Fig.~\ref{fig:figure_4}~\textbf{(c)} (upper matrix), and resulting estimated capacitances (see Appendix~\ref{sec:stability_map_simulation}). The resulting voltage cross-derivative of the ground state of the Hamiltonian, $\mathrm{d} ( \mathrm{d} E_{g} / \mathrm{d} V_{\mathrm{P3}}) / \mathrm{d} V_{\mathrm{P2}}$, is shown in Fig.~\ref{fig:figure_3}~\textbf{(d)}. See Appendix~\ref{sec:stability_map_simulation} for details of the simulation, and for the parameters used. The simulation displays qualitative agreement with data, and confirms the charge configurations $(n_{\mathrm{P2}}, n_{\mathrm{T}}, n_{\mathrm{P3}})$. The measured sensor slope in the $(n_{\mathrm{P3}}, n_{\mathrm{P2}}) = (1,1)$ is $a_{\mathrm{T}} = -0.703 \pm 0.008$, while the choice of lever arm matrix in the simulation leads to $a_{\mathrm{T}} = -0.739$. The experimental and simulated (P2,T) charge induced voltage shifts along $V_{\mathrm{P2}}$ agree within experimental resolution of $\pm 1$~mV, $\Delta V_{\mathrm{P2}} = 13 \pm 1$~mV. 
\section{Outlook}
We have used the EQD as a rf-SEB charge sensor capable of sensing QDs up to $355$~nm away from the EQD center, suggesting that the same SEB charge state may be sensitive to charges in QDs separated by over $700$~nm. Our results are well supported by quantum mechanical electron density simulations. The enhanced functionality provided by the EQD may be expanded in future QD-based architectures to sensors defined with more complex gate shapes, such as a right-angle or a cross. A single sensor could allow sensing multiple QDs placed around the periphery, enabling novel unit cells requiring fewer individual gate structures for readout. Combined with the demonstration of few-electron QDs, our results show the potential of this multi-gate polysilicon platform to produce scalable QD unit cells.
\par
Another potential application of this type of elongated QD is as a mid-range spin qubit coupler as previously demonstrated for QDs in GaAs/AlGaAs heterostructures~\cite{malinowski2019fast}. We have here demonstrated two basic requirements towards this application: the quantization of charge in the EQD and the tunnel coupling to QDs at the periphery. 
We envision that extended QDs could become an important resource to increase the range of qubit-qubit interaction in silicon, complementing other approaches such as spin shuttling ~\cite{yoneda2021coherent, noiri2022shuttling, seidler2022conveyor}, capacitive coupling with floating gates~\cite{gilbert2020single, duan2020remote} and microwave photonic links~\cite{borjans2020resonant, harveycollard2022coherent}. Additionally, we have shown that the EQD can be used as a local electron reservoir, which can be utilized in schemes mitigating charge leakage errors~\cite{cai2019silicon}.
\section{Acknowledgements}
This research was supported by European Union’s Horizon 2020 research and innovation programme under grant agreement no.\ 951852 (QLSI), and by the UK's Engineering and Physical Sciences Research Council (EPSRC) via QUES2T (EP/N015118/1), and the Hub in Quantum Computing and Simulation (EP/T001062/1). AC acknowledges funding from the Danish Independent Research Fund. M.F.G.-Z.\ is a UKRI Future Leaders Fellow (MR/V023284/1). 
\appendix
\renewcommand{\thesection}{\Alph{section}}
\renewcommand\thefigure{\thesection\arabic{figure}}
\section{Cryogenic device characterization}
\label{sec:cryogenic_device_characterization}
\begin{figure}
\includegraphics[]{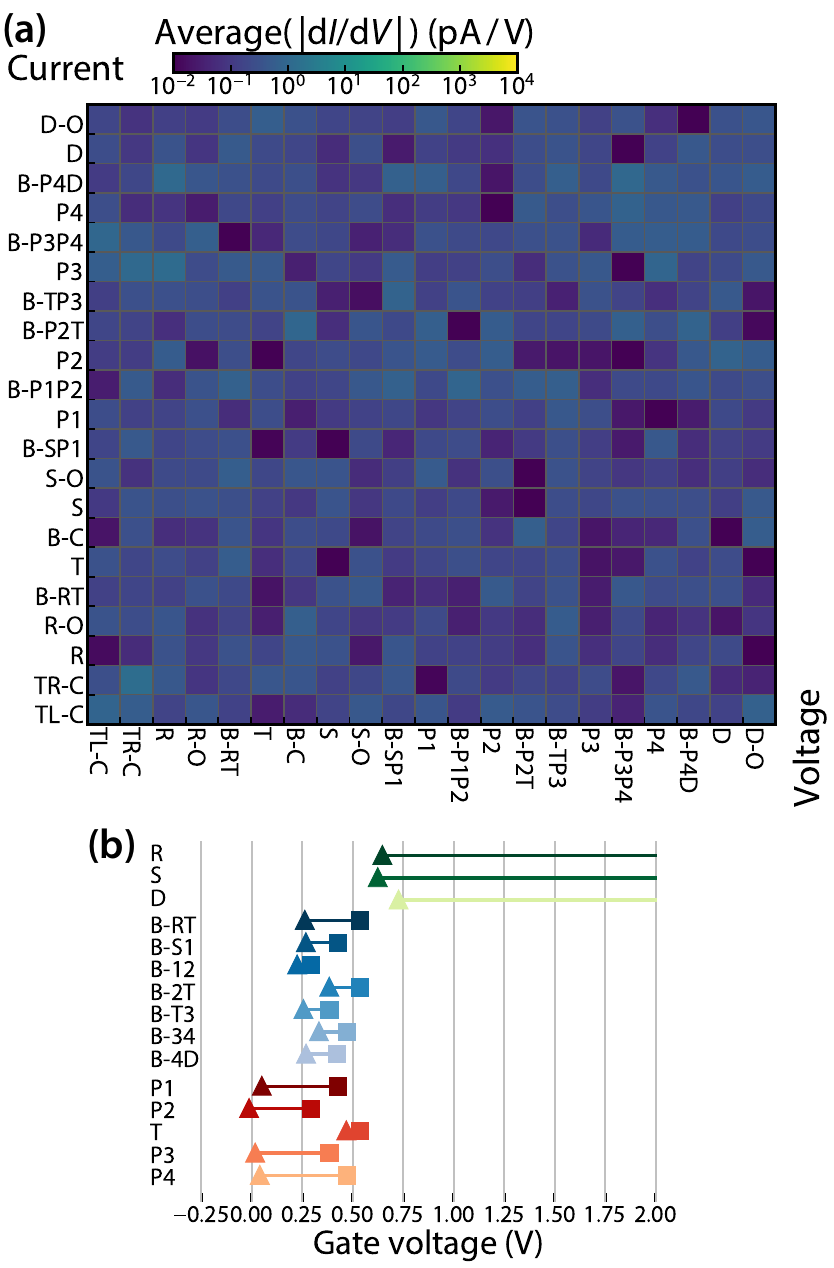}
\caption{
\label{fig:supplementary_figure_pinch_offs}
\textbf{Leakage currents, pinch-off and saturation voltages.} 
\textbf{(a)} Leakage matrix, where the colour of a pixel shows the average conductance between gate $i$ (matrix columns), where we apply a voltage, and channel $j$ (matrix rows), where we read out current. Applied voltage ranges from $-4.8$ mV to $+4.8$ mV. 
\textbf{(b)} Summary of pinch-off (triangle markers) and saturation (square markers) voltages. All gates except for B-RT and R are characterized through the source-drain channel, whereas gates B-RT and R are characterized through drain-reservoir channel. We offset the results on the y-axis and label each result with the abbreviated gate name. Red datasets correspond to layer 3 plunger gates, blue to layer 2 barrier gates, and green to layer 3 accumulation gates. 
}
\end{figure}

To assess operability of the device measured in the main text, labelled as device \textbf{A}, we measure gate leakage conductances, and pinch-off and saturation voltages at the base temperature of the cryostat. Gate leakages are measured by applying an increasing voltage to gate $g_{i}$ while measuring current through all channels, and repeating for all gate electrodes $i$. Leakage conductance $l_{ij}$ is taken as the average conductance over the voltage range. The resulting leakage matrix is shown in Fig.~\ref{fig:supplementary_figure_pinch_offs}~\textbf{(a)}. The device under study has leakage currents no larger than $\pm 1.3$ pA$/$V at cryogenic temperatures.
\par
The first guess for Coulomb blockade operating point is obtained from gate pinch-off and saturation voltage measurements, where we operate the device similarly to a circuit of classical MOS field-effect transistors in series.  
The device has three possible current channels: source-drain, reservoir-source, and drain-reservoir. Pinch-off voltages are measured by applying a $V_{\mathrm{o}i-\mathrm{o}j} = 1$~mV between ohmics $\mathrm{o}i$ and $\mathrm{o}j$ along channel $i-j$, and by biasing all layer 2 and 3 gates along said channel at $2.0$~V. The gate voltage of one of those gates is swept from $2.0$ V to $-0.5$ V and back while recording current. Pinch-off and saturation voltages are defined as the voltages where measured current is 5\% and 90\% of the saturation current, respectively. If saturation is not observed, pinch-off current is defined as 5\% of the maximum measured current. 
\par
In this device, all three channels are functional. A summary of the results, obtained by biasing the source-drain and drain-reservoir channels, is shown in Fig.~\ref{fig:supplementary_figure_pinch_offs}~\textbf{(b)}.
The accumulation gates source (S), reservoir (R), and drain (D) saturate at higher voltages than other gates. Other layer 3 gates systematically require a larger voltage range between pinch-off and saturation than layer 2 gates. This observation is consistent with the increasing total oxide thickness with increasing layer index.
Hysteresis was not observed in this device. 
\par
\section{Radiofrequency reflectometry and data acquisition} \label{sec:radiofrequency_reflectometry}
\begin{figure}
\includegraphics[width=240 bp]{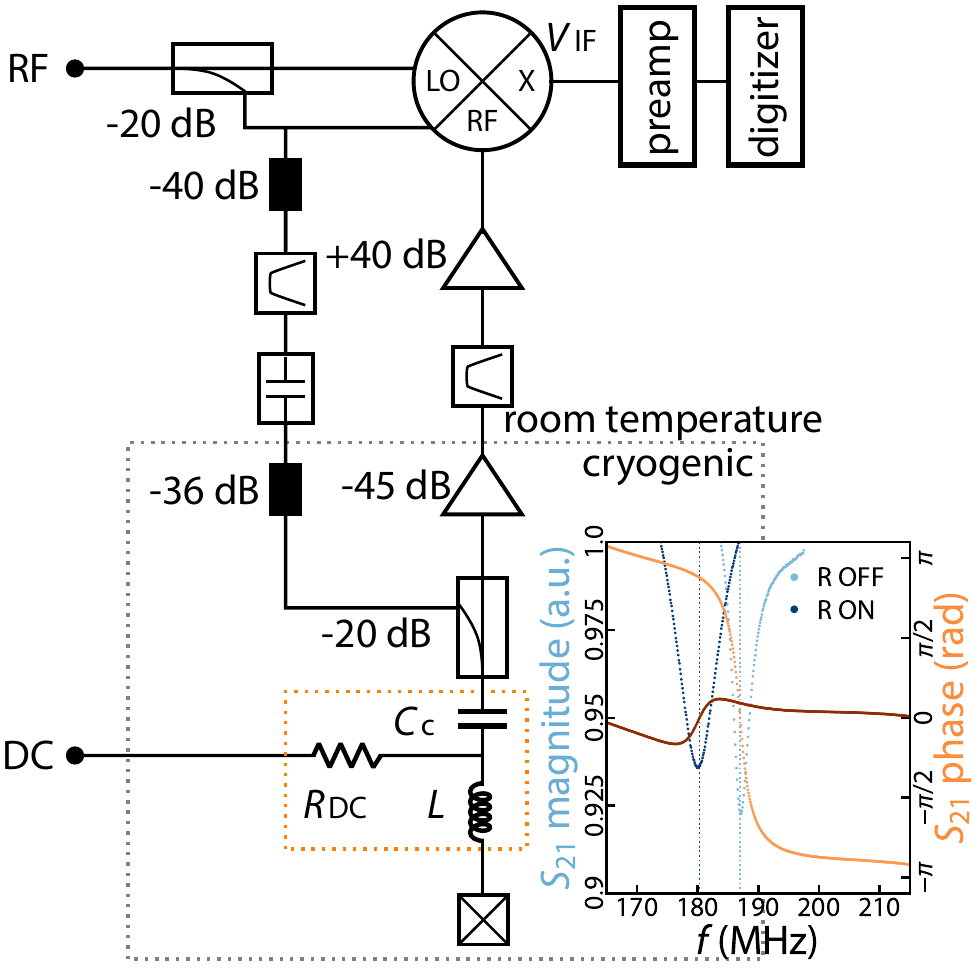}
\caption{
\label{fig:supplementary_figure_reflectometry}
\textbf{Radiofrequency reflectometry.} 
Simplified cryogenic and room-temperature rf setup, with a lumped-element resonator (circuit bordered with orange dotted rectangle) attached to the ohmic of the accumulation gate R (crossed square). Attenuation is performed in stages inside the cryostat. The total attenuation of $-36$ dB is shown for simplicity. 
DC filtering is not shown. 
Embedded plot shows the resonator response in a  vector network analyzer reflection measurement, in which the input and output signal are split for attenuation and cryogenic amplification. Signal magnitude is shown as blue datapoints up to a $6$ dB bandwidth. Signal phase is shown as orange datapoints. 
Datasets with the accumulation gate R off at zero bias are plotted as lighter datasets, and datasets with gate R on at positive bias as darker datasets. 
Estimated resonance frequencies are denoted with dotted lines. 
}
\end{figure}
\par
Figure~\ref{fig:supplementary_figure_reflectometry} shows a schematic of the RF reflectometry setup for biasing the lumped-element resonator attached to the ohmic of the accumulation gate R, and acquiring the reflected signal. The diagram includes attenuation, amplification, and the key components. 
Here, $R_{\mathrm{DC}} = 49.99$ k$\Omega$, $L = 820$ nH, and $C_{\mathrm{c}} = 22$ pF. There is also an overall coupling capacitor of $100$ pF in series with $C_{\mathrm{c}}$, omitted for simplicity. 
\par 
All device gates are connected to DC lines (not shown), which are biased with a voltage digital-to-analogue converter (DAC). To acquire a trace or a stability diagram, the DAC is programmed to send one- or two-dimensional step-discretized voltage ramps, with a step taking $t_{\mathrm{sample}} = 10^{-4}$~s corresponding to ramp frequency $f_{\mathrm{ramp}} = 10$~kHz. The start of the ramp is synchronized to a trigger sent to the digitizer. We digitize $V_{\mathrm{IF}}$ using a sample rate of $1$~MS/s, with the voltage preamplifier low-pass cutoff $f_{\mathrm{low\ pass}} = 10$~kHz. We boxcar filter the trace with a window size corresponding to $t_{\mathrm{sample}}$, and decimate accordingly to get a $t_{\mathrm{sample}}$-spaced dataset. We average by repeating acquisition $n_{\mathrm{average}} = 10$ times. Hence, we estimate the integration time $t_{\mathrm{integrate}} = n_{\mathrm{average}}/f_{\mathrm{low\ pass}} = 1$~ms. 
\par
Embedded to Fig.~\ref{fig:supplementary_figure_reflectometry} is the vector network analyzer (VNA) response with the gate R (and the entire device) at zero bias (light blue and light orange datasets for magnitude and phase responses), and with the gate R on (dark blue and dark orange datasets, respectively). The resonator responds both by changing its resonance frequency and coupling coefficient \textendash we observe the resonance frequency and the coupling coefficient to decrease, taking the resonator from overcoupled to undercoupled as a 2DEG is accumulated under gate R. 
\section{Charge sensing of quantum dots controlled with B-2T and B-T3}
\label{sec:barrier_quantum_dot_sensing}
\begin{figure*}
\includegraphics[]{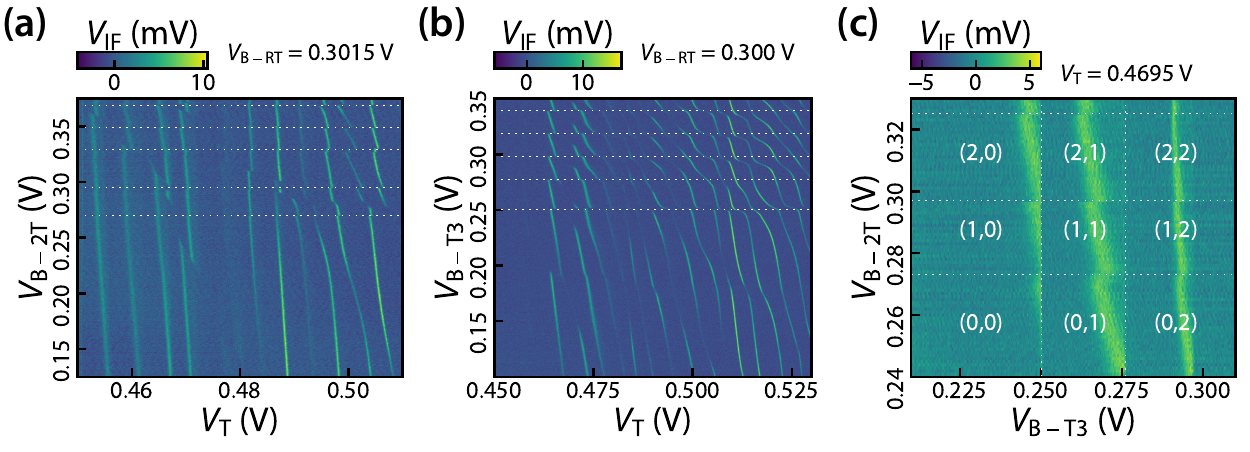}
\caption{
\label{fig:supplementary_figure_barrier_sensing}
\textbf{Charge sensing of quantum dots controlled with B-2T and B-T3.}
Stability diagrams on the \textbf{(a)} (T,B-2T), \textbf{(b)} (T,B-T3), and \textbf{(c)} (B-2T,B-T3) voltage planes. Electron loading voltages are indicated with white, dashed lines. Estimated charge configurations in \textbf{(c)} are denoted as  $(n_{\mathrm{B-2T}}, n_{\mathrm{B-T3}})$. 
}
\end{figure*}
\par
We can use the SEB to sense QDs controlled with barrier gates B-2T and B-T3. To this end, we set $V_{\mathrm{P2}} = V_{\mathrm{P3}} = 0$~V. We operate the EQD at a lower occupancy, at around $V_{\mathrm{T}} \approx 0.5$~V, compared to the $V_{\mathrm{T}} \approx 0.7$~V from main text. We estimate that the occupancy at this voltage is less by approximately $45 - 50$ electrons. We use a lower $V_{\mathrm{T}}$ to reduce the EQD electron wavefunction size to reduce the EQD-barrier dot electron tunnel coupling. 
\par
We show the (T,B-2T), (T,B-T3), and (B-2T,B-T3) stability diagrams in Figure~\ref{fig:supplementary_figure_barrier_sensing} \textbf{(a)}-\textbf{(c)}. 
In this $V_{\mathrm{T}}$ range, we observe DRTs which couple more strongly to the T gate than the barrier gates. It is possible that in this regime the gate T hosts multiple QDs.
Similarly to when sensing the plunger QDs (see Fig.~\ref{fig:figure_2}), we observe the first few electrons being loaded under B-2T and B-T3 as shifts in the EQD DRT peaks. We observe approximately monotonic increase in tunnel coupling in the (T,B-T3) system (Fig.~\ref{fig:supplementary_figure_barrier_sensing} \textbf{(b)}). 
The barrier-QD to EQD coupling changes from capacitively coupled to tunnel coupled, as indicated by the increasing bending of the DTR lines for increasing $V_{\mathrm{B-T3}}$ and $V_{\mathrm{T}}$ voltages, 
over a range of $\Delta V_{\mathrm{T}} = 0.1$ V, or a change in T occupancy by approximately 15 electrons. 
We expect this to correspond to an increase in the EQD electron wavefunction overlap with the B-T3 QD. 
\par
We also observe B-2T electron loading events in the (T,B-2T) stability diagram (Fig.~\ref{fig:supplementary_figure_barrier_sensing} \textbf{(a)}). Here, capacitive coupling is weaker at $V_{\mathrm{T}} \leq 0.49$ V, below which the B-2T addition voltages are not visible. 
This asymmetry is directly demonstrated on the (B-2T, B-T3) stability diagram, shown in  Fig.~\ref{fig:supplementary_figure_barrier_sensing} \textbf{(c)}. The EQD DRTs appear almost vertical, indicating relatively stronger coupling to B-T3 compared to B-2T. 
Based on the experimentally observed ratio 
$\alpha_{\mathrm{T},\mathrm{B-2T}}/\alpha_{\mathrm{T},\mathrm{B-T3}}$ (or 
$\alpha_{\mathrm{T},\mathrm{P3}}/\alpha_{\mathrm{T},\mathrm{P2}}$), we expect that the electron wavefunction shape and delocalization depends on filling. At low dot occupancy, the wavefunctions are smaller and not centered, with at least two possible causes. The opening between top left and top right confinement gates is asymmetric with respect to the gate T, while the L-shape of B-RT asymmetrically screens the gate T.
An additional effect may be played by orbital angular momentum. In analogy to atomic $p$ orbitals, for example, electron wavefunctions corresponding to different fillings might be centered towards the left or right of the gate T region.
For higher filling, where the EQD behaves as a single QD, the wavefunction shapes homogenize and delocalize, capacitively coupling to the outer QDs and thereby forming a well-defined, nearly symmetrically coupled TQD (see main text).
\section{Elongated quantum dot charging energies and lever arms}
\label{sec:T_charging_energies}
\begin{figure}
\includegraphics[]{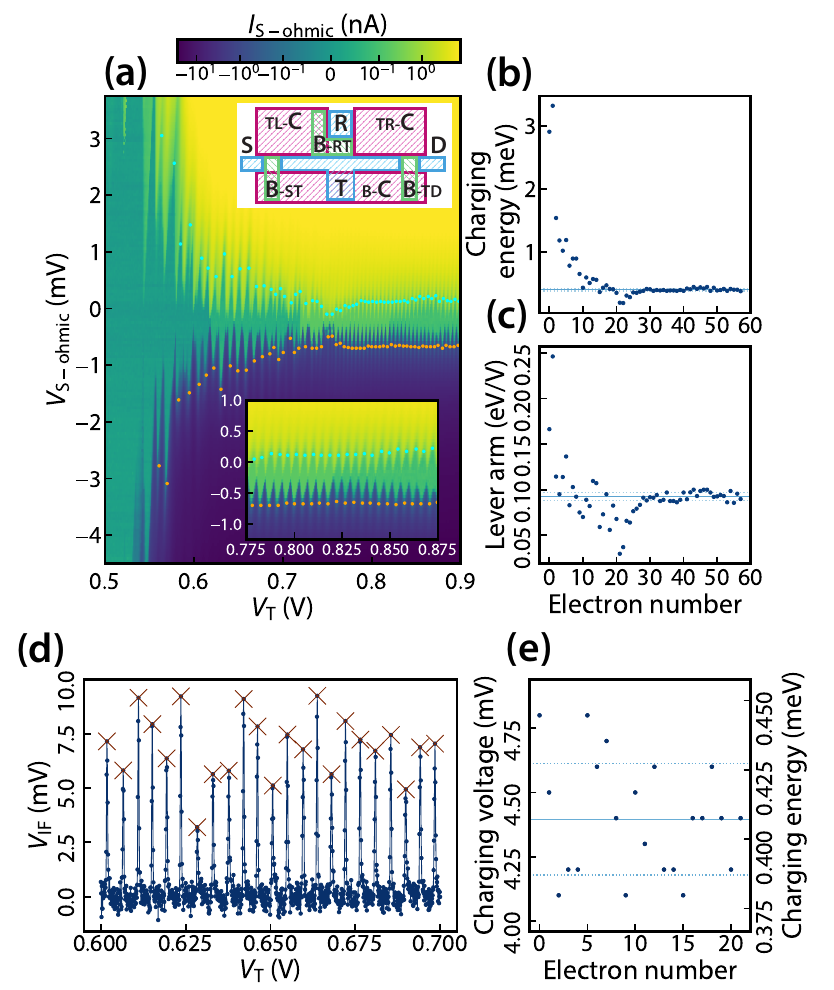}
\caption{
\label{fig:supplementary_figure_Tgate_charging_energies}
\textbf{Elongated quantum dot charging energies and lever arms.}
\textbf{(a)} Coulomb diamonds of device \textbf{B}, with diamond peaks and dips annotated as cyan and orange dots. Top inset shows the device schematic. Bottom inset shows a closeup in the high-electron-number regime. 
\textbf{(b)} Extracted charging energies for 58 electrons, starting from some offset number of electrons. 
\textbf{(c)} Extracted lever arms. 
\textbf{(d)}-\textbf{(e)} Loading and charging voltages of device \textbf{A}. Panel \textbf{(e)} shows charging voltages converted to charging energies using $\alpha_{\mathrm{T},\mathrm{T}}$ from device \textbf{B}. 
}
\end{figure}
\par
Coulomb diamond data, shown 
in Fig.~\ref{fig:supplementary_figure_Tgate_charging_energies} \textbf{(a)}-\textbf{(c)}, from another elongated quantum dot device, referred to as device \textbf{B}, is used to estimate the typical EQD charging energies and lever arms $\alpha_{\mathrm{T}, \mathrm{T}}$. The device schematic is shown in the top inset of  Fig.~\ref{fig:supplementary_figure_Tgate_charging_energies} \textbf{(a)}: the EQD is connected to three accumulation gates, S, R, and D, via barrier gates B-ST, B-RT, and B-TD. The device is from the same die as device \textbf{A}, and the EQD dimensions are the same. 
\par
We measure Coulomb diamonds through the S-D channel by applying a varying voltage to the ohmic connected to gate S, and a varying voltage to gate T. Barriers B-ST and B-TD are biased close to their pinch-off voltages, while the accumulation gates S and D are biased significantly above their respective pinch-off voltages. 
We plot the Coulomb diamonds in Fig.~\ref{fig:supplementary_figure_Tgate_charging_energies} \textbf{(a)}. For each $V_{\mathrm{T}}$, we find the diamond edges as the $V_{\mathrm{S-ohmic}}$ values where current is $\pm$ 400 pA above the background. Treating upper and lower diamond edges as data traces with setpoints given by $V_{\mathrm{T}}$, we find peak maxima and minima, which are denoted with cyan and orange points. We manually remove some of the unpaired peaks and dips from the low-electron regime. Ensuring we have an equal number of diamond peaks and dips, we process these to charging energies, shown in Fig.~\ref{fig:supplementary_figure_Tgate_charging_energies} \textbf{(b)}, and lever arms, shown in Fig.~\ref{fig:supplementary_figure_Tgate_charging_energies} \textbf{(c)}, using $E_{C i} = h_{i}/2$, and $\alpha_{\mathrm{T} i,\mathrm{T}} = h_{i}/ (2 w_i)$, where $h_{i}$ and $w_{i}$ are the height and width of $i$th diamond, respectively. 
\par
We find that the charging energies and lever arms settle to constant value for the last $30$ or so measured electrons (voltages $V_{\mathrm{T}} = 0.77$ V and above; see the bottom inset of Fig.~\ref{fig:supplementary_figure_Tgate_charging_energies} \textbf{(a)} for a close-up of the data). The average charging energy for the last $30$ electrons is $E_{C} = 0.41 \pm 0.02$~meV, and the average lever arm $\alpha_{\mathrm{T},\mathrm{T}} = 0.093 \pm 0.005$ eV/V, where error bars are given for one standard deviation. 
\par
n Fig.~\ref{fig:supplementary_figure_Tgate_charging_energies} \textbf{(d)}, we plot the Coulomb oscillations of device \textbf{A} measured in rf and extract the position of each peak to determine the charging voltages. In Fig.~\ref{fig:supplementary_figure_Tgate_charging_energies} \textbf{(e)}, we plot the extracted charging voltages, and charging energies by utilising the lever arm estimated for device \textbf{B}.
Using the lever arm estimated for the standalone gate T device, we obtain the average charging energy $E_{C} = 0.41 \pm 0.02$ meV (see extra y-axis in Fig.~\ref{fig:supplementary_figure_Tgate_charging_energies} \textbf{(f)}), which agrees well with the device \textbf{B}.
\section{Lever arm estimation}
\label{sec:lever_arm_estimation}
\begin{figure*}
\includegraphics[]{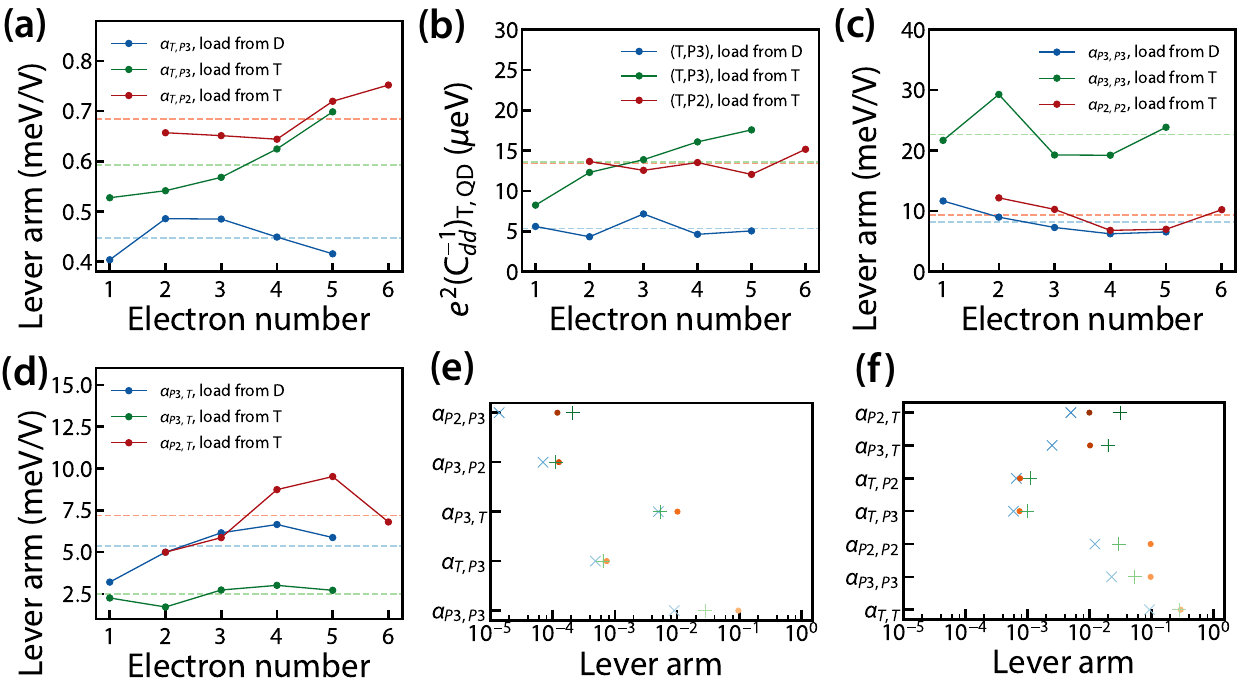}
\caption{
\label{fig:supplementary_figure_lever_arm_estimation}
\textbf{Experimentally estimated lever arms.}
Connected datapoints show the estimated 
\textbf{(a)} $\alpha_{\mathrm{T},\mathrm{P2}}$ and $\alpha_{\mathrm{T},\mathrm{P3}}$, 
\textbf{(b)} mutual charging energies between the T QD and the P3 or P2 QDs, 
\textbf{(c)}  $\alpha_{\mathrm{P2},\mathrm{P2}}$ and $\alpha_{\mathrm{P3},\mathrm{P3}}$,  
and \textbf{(d)}  $\alpha_{\mathrm{P2},\mathrm{T}}$ and $\alpha_{\mathrm{P3},\mathrm{T}}$. 
Dashed lines show averages over electron number. Legends further specify the operating point. 
\textbf{(e)}-\textbf{(f)} Comparison between experimentally estimated and simulated lever arms. Panel \textbf{(e)} shows the lever arms from operating point where the B-T3 is off at zero bias, and panel \textbf{(f)} obtained with B-T3 on at non-zero bias.  
Experimentally estimated lever arm components are drawn with blue diagonal cross markers (x). Lever arms simulated in a fixed operating point, biasing each QD-controlling gate, and without any other gate biases, are drawn with red filled circles. Lever arms simulated with biases at QD controlling gates, and nearest-neighbour gates, are drawn with green plus cross markers ($+$). 
}
\end{figure*}

The lever arm matrix is defined as the product 
\begin{align}
\pmb{\alpha} = - \big( \mathrm{\textbf{C}}_{\mathrm{dd}} \big)^{-1} \mathrm{\textbf{C}}_{\mathrm{dg}},
\label{eq:lever_arm_matrix_definition}
\end{align}
where $\mathrm{\textbf{C}}_{\mathrm{dd}}$ is the dot-dot and $\mathrm{\textbf{C}}_{\mathrm{dg}}$ the dot-gate submatrix of the Maxwell capacitance matrix~\cite{mills2019shuttling}. 
We use the device \textbf{B} high electron number average estimate $\alpha_{\mathrm{T},\mathrm{T}} = 0.093 \pm 0.005$ eV/V to convert lever arm ratios measured in device \textbf{A} to lever arms. 
The slope of a dot-to-reservoir transition on the $(V_{x}, V_{y})$ voltage plane is given by 
\begin{align}
a_{i} 
&= 
- \frac{ \alpha_{ix} }{ \alpha_{iy} }.
\end{align}
For $i = \mathrm{T}$, $x = \mathrm{P2}$ ($x = \mathrm{P3}$), and $y = \mathrm{T}$, slope estimation with $\alpha_{\mathrm{T},\mathrm{T}}$ yields an estimate for $\alpha_{\mathrm{T},\mathrm{P2}}$ ($\alpha_{\mathrm{T},\mathrm{P3}}$). Using data from Fig.~\ref{fig:figure_2}, we obtain the lever arm estimates plotted in Fig.~\ref{fig:supplementary_figure_lever_arm_estimation}~\textbf{(a)}. Likewise, the inter-site charging energies $e^2 \big( \mathrm{\textbf{C}}_{dd}^{-1} \big)_{ij}$ 
can be estimated from the horizontal and vertical ICT extents $\Delta V_{x}$ and $\Delta V_{y}$, as
\begin{align}
\Delta V_{x}
&=
-2\big(\mathrm{\textbf{C}}_{dd}^{-1}\big)_{ij}
\frac{
\alpha_{iy} - \alpha_{jy}
}
{
\alpha_{ix} \alpha_{jy} - \alpha_{iy} \alpha_{jx}
}
\label{eq:horizontal_ICT_extent}
\\
\Delta V_{y}
&=
a_{ij} (x_{23} - x_{12}),
\label{eq:vertical_ICT_extent}
\end{align}
for dots $i$ and $j$, where $a_{ij}$ is the slope of the ICT. 
When $\alpha_{\mathrm{P2},\mathrm{}P2} \gg \alpha_{\mathrm{T},\mathrm{P2}}$, Eq.~\eqref{eq:horizontal_ICT_extent} simplifies to 
\begin{align}
\Delta V_{\mathrm{T}}
&=
2 |e| \big( \mathrm{\textbf{C}}_{dd}^{-1}
\big)_{\mathrm{T},\mathrm{P2}}
\alpha_{\mathrm{T},\mathrm{T}}^{-1},
\label{eq:approximate_ICT_extent_along_VT}
\end{align} 
and when $\alpha_{\mathrm{T},\mathrm{T}} \gg \alpha_{\mathrm{P2},\mathrm{T}}$, Eq.~\eqref{eq:vertical_ICT_extent} simplifies to 
\begin{align}
\Delta V_{\mathrm{P2}}
&\approx
- 2 |e| \big( \mathrm{\textbf{C}}_{dd}^{-1} \big)_{\mathrm{T},\mathrm{P2}} 
\alpha_{\mathrm{P2},\mathrm{P2}}^{-1},
\label{eq:approximate_ICT_extent_along_VP2}
\end{align}
and similarly for $\mathrm{P3}$. 
\par
The estimates obtained for mutual capacitances $\big( \mathrm{\textbf{C}}_{dd}^{-1}
\big)_{\mathrm{T},\mathrm{P2}}$ and $\big( \mathrm{\textbf{C}}_{dd}^{-1}
\big)_{\mathrm{T},\mathrm{P3}}$, and the lever arms $\alpha_{\mathrm{P2},\mathrm{P2}}$ and $\alpha_{\mathrm{P3},\mathrm{P3}}$ 
by inverting Eqs.~\eqref{eq:approximate_ICT_extent_along_VT}-\eqref{eq:approximate_ICT_extent_along_VP2} 
are plotted in Fig.~\ref{fig:supplementary_figure_lever_arm_estimation}~\textbf{(b)}-\textbf{(c)}.
Using the above estimates, we may also estimate $\alpha_{\mathrm{P2},\mathrm{T}}$ and $\alpha_{\mathrm{P3},\mathrm{T}}$ by inverting the ICT slope 
\begin{align}
a_{\mathrm{P2},\mathrm{T}}
&:=
\frac{
\Delta V_{\mathrm{T}}
}{
\Delta V_{\mathrm{P2}}
}
\\
&=
\frac{ \alpha_{\mathrm{P2},\mathrm{P2}} - \alpha_{\mathrm{T},\mathrm{P2}} 
}{
\alpha_{\mathrm{P2},\mathrm{T}} - \alpha_{\mathrm{T},\mathrm{T}}
}
.
\end{align}
The resulting estimates are plotted in Fig.~\ref{fig:supplementary_figure_lever_arm_estimation}~\textbf{(d)}.
\par
We point out that the setpoint for data in Fig.~\ref{fig:figure_3} \textbf{(c)} is locally close to the setpoints for the (P2,T) and (T,P3) DQDs discussed in Figs.~\ref{fig:figure_2} \textbf{(a)} and \textbf{(c)}, where B-T3 is held at zero bias. This is demonstrated by an independent lever arm ratio estimate. For the charge configurations $(n_{\mathrm{P2}}, n_{\mathrm{T}}) = (1, n_{\mathrm{T}})$ and $(n_{\mathrm{T}}, n_{\mathrm{P3}}) = (n_{\mathrm{T}}, 1)$, 
we extract 
$\alpha_{\mathrm{T},\mathrm{P2}}/\alpha_{\mathrm{T},\mathrm{T}} = 0.140 \pm 0.008$ 
and 
$\alpha_{\mathrm{T},\mathrm{P3}}/\alpha_{\mathrm{T},\mathrm{T}} = 0.097 \pm 0.016$, 
yielding 
$\alpha_{\mathrm{T},\mathrm{P3}}/\alpha_{\mathrm{T},\mathrm{P2}} = 0.69 \pm 0.14$, agreeing with the previous estimation. 
Motivated by this observation, in the following, when simulating the TQD stability diagram of Fig.~\ref{fig:figure_3} \textbf{(c)}, we use lever arms and inter-site charging energies estimated from these DQD datasets. 
\par
The above procedure yields estimates for $7$ out of the $9$ lever arms for the (P2,T,P3) TQD system. The diagonal lever arms $\alpha_{\mathrm{P2},\mathrm{P2}}$ and $\alpha_{\mathrm{P3},\mathrm{P3}}$ also enable to convert addition voltages to addition energies. In principle, we may estimate the remaining two from the P3 and P2 DRTs on the (P3,P2) stability diagram. The high resolution along $V_{\mathrm{P3}}$ enables to estimate the slope $-\alpha_{\mathrm{P3},\mathrm{P3}}/\alpha_{\mathrm{P3},\mathrm{P2}}$, whereas the lower resolution along $V_{\mathrm{P2}}$ renders estimating $-\alpha_{\mathrm{P2},\mathrm{P3}}/\alpha_{\mathrm{P2},\mathrm{P2}}$ more difficult. 
We estimate the latter in a charge configuration of $3...6$ electrons under P2 and P3, in $V_{\mathrm{P2}}$ and $V_{\mathrm{P3}}$ range of $0.6 ... 0.9$ V.
\par
\section{Self-consistent Schr{\"o}dinger-Poisson and electrostatic solvers}
\label{sec:self_consistent_schroedinger_poisson_simulation}
The quantum-mechanical electron density (QMED) $n^{-}(\textbf{r})$ is defined, as the energy integral of a sum of Fermi-distributed probability densities corresponding to different conduction bands~\cite{birner2011modeling}.
The QMED can be obtained by iteratively solving the Schr{\"o}dinger and Poisson equations~\cite{nextnanoWebManual, birner2011modeling}, method referred to as self-consistent Schr{\"o}dinger-Poisson solver (SPS). At convergent energies, the solver outputs both the QMED $n^{-}(\textbf{r})$, and the so-called effective-mass probability density $\big| \Psi_{\alpha, E}(\textbf{r}) \big|^{2}$. Total charge associated with an electron density is obtained by integrating $n^{-}(\textbf{r})$ over space. 
\par
We use an SPS implementation from the semiconductor nanostructure simulation program nextnano++~\cite{nextnanoWebManual, birner2011modeling}. We model the device gates as 3D Schottky contacts over a silicon (Si) substrate. We take the work-function $\phi = 4.05$ eV, consistent with $n^{++}$-doped polycrystalline silicon. We take the Si/SiO$_2$ interface as a Dirichlet boundary condition, which enforces the electron densities to remain within the Si substrate. We also employ the electrostatics module from the general-purpose simulation software COMSOL multiphysics, to evaluate the Maxwell capacitance matrix of a system of metallic objects. 
\par
We draw the device gate model in COMSOL and nextnano++ layer by layer, using device gate and oxide dimensions consistent with device design and expected fabricated dimensions. 
Gate layer colouring follows the device schematic of Fig.~\ref{fig:supplementary_figure_reflectometry} \textbf{(a)}.  At gate overlaps, 
we draw a gate in the overlapping area, in general creating several, not necessarily connected, objects to describe a single gate. We define higher- and lower-level gate objects as a single electrical node.
\par
To estimate the shape of a QD that is controlled with gate $A$ in an experiment, we bias the gate $A$ and it's nearest-neighbour gates in the simulation according to the experimental setpoint, while retaining all other gate biases at zero volts. In these simulated setpoints, we use the SPS to solve for the electron and probability densities. Thus, the simulation has two differences compared to experiments. In the simulation, we do not employ the reservoir gates S, R, and D; and only a subset of the device is biased in one simulation. We split the QD QMED simulation into multiple parts as opposed to simulating all QMEDs in a single simulation for two reasons. 
High electron densities, such as those due to electron reservoirs S, R, and D, converge more slowly, significantly incresing runtime. Also, in our experiment, we operate plunger gate controlled QDs in the few electron regime, while we operate the EQD at an occupancy of $\approx 50-60$ electrons. We expect the reservoir electron densities to be approximately a factor of $100$ higher than the QD electron densities. Due to the large difference in magnitude, the convergence of the SPS solver depends mostly of the reservoir electron density, leaving the estimate for the QD electron density inaccurate (e.g. simply void of electrons), and significantly different than the estimate obtained without including the reservoir. 
\par
We assimilate the QMED with the wavefunction, as $\rho(\textbf{r}) = n^{-}(\textbf{r})$. We prefer electron densities instead of probability densities to describe the many-electron states in the EQD. Alternatively, we could evaluate probability densities up to the number of electrons we expect at the EQD. Probability densities have orbitals with more irregularity in shape and size as a function of electron number. These states do not account for electron-electron interactions, and hence we do not expect a gain in accuracy by using probability densities. We then consistently employ electron densities for all QD shape estimates.  
\par
Shapes of a QD at e.g. $n \sigma$ are taken as the 3d contour $\textbf{r}_{\mathrm{boundary}}$ at which the density has fallen by $n \sigma$ from the maximum. That is, 
\begin{align}
\textbf{r}_{\mathrm{boundary}}
&=
\big\{ \textbf{r} : \rho(\textbf{r}) = (1 - n \sigma) \rho_{\mathrm{max}}(\textbf{r}) \big\}.
\label{eq:electron_density_boundary}
\end{align}
\section{Simulated capacitance matrices}
\label{sec:simulated_capacitance_matrices}
To estimate the Maxwell capacitance matrix $\mathrm{\textbf{C}}$, the estimated QD boundaries (Eq.~\eqref{eq:electron_density_boundary})are imported to COMSOL as shapes, and defined as a perfect hollow conductor which is maintained at zero bias. We use the $1 \sigma$ contours in all our capacitance matrix simulations. Having imported all QD shapes to the same device model, we run the electrostatic solver to obtain $\mathrm{\textbf{C}}$. The lever arm matrix is obtained from Eq.~\eqref{eq:lever_arm_matrix_definition}. 
\par
As a reference dataset, we calculate a capacitance matrix for the TQD system (P2,T,P3), where QD electron densities are obtained in a simulation where only the corresponding plunger gate is biased. We compare this to a simulation, where we account for the effect that barriers have for QD shapes and locations, by biasing a plunger gate and nearest neighbour barrier gates. We choose the setpoints corresponding to B-T3 on and B-T3 off (see Figs.~\ref{fig:figure_2} and \ref{fig:supplementary_figure_reflectometry}). 
\par
We plot the simulated lever arms corresponding to the no-barrier and B-T3 off, as well as the experimentally estimated B-T3 off lever arms, in Fig.~\ref{fig:supplementary_figure_lever_arm_estimation} \textbf{(e)}. 
The simulated and experimental lever arms corresponding to no-barrier and B-T3 on are plotted in Fig.~\ref{fig:supplementary_figure_lever_arm_estimation} \textbf{(f)}. We have averaged over charge configurations for the experimental lever arms. We find that the simulated lever arms are systematically larger than experimentally estimated lever arms. Simulations with and without the neighbouring barriers produce estimates with similar magnitudes, but simulations with barriers qualitatively follow the trends observed in the experimental lever arms. We associate the systematic mismatch between experimental and simulated lever arm components to the fact that the simulation systematically overestimates charging energies $|e| \big( \mathrm{\textbf{C}}_{\mathrm{dd}}^{-1} \big)_{ij}$, i.e. underestimates the dot-dot capacitances. One source of error is our inability to describe the large electron reservoirs, to which the EQD, and at some setpoints the P3 QD, are highly tunnel coupled to. The tunnel coupling affects the QD shape.  
\par
In the main text, we summarize these results using the relative error matrix between experimental, $\alpha_{\mathrm{exp},ij}$, and simulated, $\alpha_{\mathrm{sim},ij}$, lever arms components, defined as 
\begin{align}
\big( \delta \alpha \big)_{ij} = |\alpha_{\mathrm{exp},ij} - \alpha_{\mathrm{sim},ij}|/| \alpha_{\mathrm{exp},ij} |
\end{align}
(see the lower matrix of Fig.~\ref{fig:figure_4} \textbf{(c)}). 
\section{Capacitively coupled triple quantum dot stability diagram simulation}
\label{sec:stability_map_simulation}
We simulate the charge stability diagram of a triple quantum dot as the ground state of the electrostatic Hamiltonian
\begin{align}
H_{\mathrm{\textbf{C}}}
&=
\sum_{i \in d} \frac{1}{2} 
\bigg[ \sum_{j \in d} e^{2} n_{i} (\mathrm{\textbf{C}}_{dd}^{-1})_{ij} \, n_{j}
+
\sum_{k \in g} e \, n_{i} \, \pmb{\alpha}_{ik} V_{k} \bigg],
\end{align}
where $e$ is the electron charge, $n_{i}$ the number operator for site $i$, $\Delta_{i}$ is the site $i$ orbital energy, $\big( \mathrm{\textbf{C}}_{\mathrm{dd}} \big)^{-1}$ the inverse of the dot-dot submatrix of the capacitance matrix, $\pmb{\alpha}$ the lever arm matrix, and $V_{k}$ the gate voltage of gate $k$. We represent the number operators using fermionic ladder operators, in term using the Jordan-Wigner mapping. We reduce to considering a single orbital per site (i.e. up to 2 electrons per site), and, since $n_{i}$ are diagonal, it suffices to consider $H_{\mathrm{\textbf{C}}}$ as a vector. 
The ground state is found as the smallest element of the vector. In order to obtain the charge stability diagram as a function of two voltages, we perform the simulation for a matrix of voltage pairs $V_{P2}, V_{P3}$. We define the stability diagram as the cross-derivative $\mathrm{d} \big[ \mathrm{d} E_{g}(V_{P2}, V_{P3}) / \mathrm{d} V_{P3} \big] / \mathrm{d} V_{P2}$, where $E_{g}(V_{P2}, V_{P3})$ is the resulting ground state matrix as a function of the two voltages.
\par
In the following, matrices are expressed in a basis with indices $\{ 1,2,3 \} = \{ \mathrm{P2}, \mathrm{T}, \mathrm{P3} \}$. 
We use the lever arm matrix (expressed in units of eV/V), which reads up to four significant figures
\begin{align}
\pmb{\alpha}
&=
\begin{pmatrix}
12.20 & 4.99 & 0.01 \\
0.66  & 92.60 &  0.49  \\
0.07 & 4.99 & 9.01 
\end{pmatrix} \times 10^{-3}.
\end{align}
The exact values are given by the estimated lever arm components, which were not truncated.
\par
We use the inverse dot-dot capacitance matrix, whose components are proportional to charging energies, expressed in units of eV, 
\begin{align}
|e| \big( \mathrm{\textbf{C}}_{\mathrm{dd}} \big)^{-1}
&=
\begin{pmatrix}
0.463  & 0.014 & 10^{-5} \\
0.014, &  0.407   & 0.004 \\
10^{-5} & 0.004  & 0.697  \\
\end{pmatrix} \times 10^{-3}.
\end{align}
The matrix is assumed to be symmetric. We assimilate addition energies to charging energies in this simulation. 
The elements $|e| \big( \mathrm{\textbf{C}}_{\mathrm{dd}} \big)^{-1}_{P2,P2}$ and $|e| \big( \mathrm{\textbf{C}}_{\mathrm{dd}} \big)^{-1}_{P3,P3}$ are estimated from the addition energies $\Delta V_{P2} (n \to n+1)$ and $\Delta V_{P3} (n \to n+1)$ of Fig.~\ref{fig:figure_3} \textbf{(c)}, as
\begin{align}
|e| \big( \mathrm{\textbf{C}}_{\mathrm{dd}} \big)^{-1}_{P2,P2}
&=
\frac{ 1 }{ 2 } \alpha_{P2,P2} \Delta V_{P2}(1 \to 2)
,
\end{align} 
and likewise for P3. 
The elements $|e| \big( \mathrm{\textbf{C}}_{\mathrm{dd}} \big)_{P2,P3}$ and its transpose are not estimated, but a small non-zero element is added to aid matrix inversion. Matrix inversion is used when converting the charging energies to a dot-dot capacitance matrix. 
\par
To account for finite threshold voltages, we shift plotting ranges for $V_{P3}$ and $V_{P2}$ after the simulation. The simulated ranges are from $-0.05$ to $+0.25$ V. We use a fixed applied voltage parameter $V_{T} = 0.0039$ V to set the T QD DRT alignment with respect to the P2 and P3 DRTs. 
The resulting simulation is shown in Fig.~\ref{fig:figure_4} \textbf{(d)}. The charge configurations are evaluated as number operator expectation values at each $(V_{\mathrm{P2}}, V_{\mathrm{P3}})$ pair, with $n_{\mathrm{T}}$ added for the middle QD charge configuration. 
\bibliography{references.bib}
\end{document}